\documentclass[prl,aps,showpacs,twocolumn,floatfix]{revtex4}
\usepackage{graphicx}

 \def\ba#1{\begin{array}{#1}}
 \def\ea{\end{array}}
 \def\be{\begin{equation}}
 \def\ee{\end{equation}}
\def\bq{\begin{equation}}
\def\eq{\end{equation}}
 \def\br{\begin{eqnarray}}
 \def\er{\end{eqnarray}}

\begin{document}
\title{  Shell-model descriptions of mass 16-19 nuclei with \\
chiral two- and three-nucleon interactions}
\author{Huan\ Dong$^{1}$, T.\ T.\ S.\ Kuo$^{1}$ and J.\ W.\ Holt$^{2}$}
\affiliation{$^1$Department of Physics, State University New York at Stony 
Brook\\ Stony Brook, New York 11794}
\affiliation{$^2$Physik Department, Technische Universit\"at M\"unchen,
D-85747 Garching, Germany}
\date{\today}

\begin{abstract}
  Shell-model calculations for several mass 16-19 nuclei
are performed using  the N$^3$LO two-nucleon potential 
$V_{2N}$ with and without
the addition of an in-medium  three-nucleon potential 
$V_{3N}^{med}$, which is a  density-dependent effective  
two-nucleon potential recently derived from the leading-order 
chiral three-nucleon force $V_{3N}$ by Holt, Kaiser, and Weise. 
We first calculate the $V_{low-k}$ low-momentum interactions 
 from $V_{2N}$ and $V_{3N}^{med}$.
The  shell-model effective interactions  for both the $sd$ one-shell
and $sdpf$ two-shell model spaces are then obtained from
these low-momentum interactions using respectively the Lee-Suzuki 
and the recently developed Okamoto and Suzuki iteration methods.
 The effects of $V_{3N}^{med}$ to the low-lying states of 
$^{18}O$, $^{18}F$, $^{19}O$ and $^{19}F$ are generally small and attractive,
mainly lowering the ground-state energies of these nuclei and making 
them in better agreements with experiments than those 
calculated with $V_{2N}$ alone. The excitation spectra of these nuclei
are not significantly affected by $V_{3N}^{med}$.
The low-lying spectra of these nuclei calculated with the $sd$ and $sdpf$
model spaces are closely similar to each other. 
Our shell-model calculations for $^{16}O$  indicate
that the $V_{3N}^{med}$ interaction is important and desirable 
for the binding energy of this nucleus.

\end{abstract}


\maketitle

\section{I. Introduction}

 The nuclear shell model has been very successful in microscopic 
descriptions of nuclear structure, and in this approach
 the  shell-model effective interaction $V_{eff}$   
has played an important role and its determination 
has been extensively studied  
\cite{brownwild88,brownrich06,jensen95,coraggio09}. As discussed in these 
references, $V_{eff}$ may be determined using either an empirical approach
where it is required to reproduce  selected experimantal data or 
a microscopic one where $V_{eff}$
is derived  from realistic nucleon-nucleon
 interactions using many-body methods. The  interactions used in
 such microscopic calculations have been mostly the two-nucleon (NN)
interaction $V_{2N}$ \cite{jensen95,coraggio09}. 
Should $V_{eff}$ have also contributions from the three-nucleon (NNN) force
$V_{3N}$ in addition to the two-nucleon one?
 In fact the need of  $V_{3N}$
in nuclear many-body problems has long been recognized. 
The use of $V_{2N}$ alone has been inadequate
in reproducing the empirical nuclear matter saturation properties
(see e.g. \cite{siu09,dong09} and references quoted therein).
The inclusion of $V_{3N}$ has been of essential importance in describing
the binding energies and low-lying spectra of light nuclie
\cite{nogga00,pieper02,navratil07} and in explaining the long half-life
of the $^{14}{\rm C} \rightarrow^{14}{\rm N}$ $\beta$-decay 
\cite{holt09,holt10}. Otsuka {\it et al.} \cite{otsuka10} 
have shown that the inclusion of $V_{3N}$
plays a crucial role in describing the oxygen isotopes near the drip
line.

In the present work, we shall calculate the shell-model effective 
interactions for the $sd$ and $sdpf$ shells using the chiral
$N^3LO$ two-nucleon potential $V_{2N}$ \cite{idaho}
with and without the inclusion of the  in-medium
NNN force $V_{3N}^{med}$, which is a density-dependent
two-nucleon potential recently derived from the leading-order
chiral NNN force $V_{3N}$ by Holt, Kaiser and Weise 
\cite{holt09,holt10}. We shall apply
these effective interactions to shell-model calculations for
 nuclei $^{18}O$, $^{18}F$, $^{19}O$ and $^{19}F$,  to study 
the effects of  $V_{3N}^{med}$ in these nuclei.
We shall first calculate the low-momentum 
$V_{low-k}$ interactions 
\cite{bogner01,bogner02,bogner03} from $V_{2N}$ and $V_{3N}^{med}$.
Our shell model effective interactions $V_{eff}$ will then be calculated
from these  $V_{low-k}$ interactions 
  using a folded-diagram formalism
\cite{jensen95,coraggio09,ko90,klr}. In this formalism, $V_{eff}$ is
given as a folded-diagram expansion. For the degenerate $sd$ case,
this expansion  can be summed up
using the commonly employed  Lee-Suzuki ({\rm LS})
iteration method \cite{lesu,sule}. For the non-degenerate $sdpf$ case
we shall sum up the expansion using the extended Krenciglowa-Kuo 
iteration method recently developed by
 Okamoto, Suzuki, Kumagai and Fujii ({\rm EKKO}) 
\cite{okamoto10,dongzbox11}. The EKKO method is
efficient for deriving the effective interactions of non-degenerate
model spaces such as the $sdpf$ two-shell space, 
and it can also be conveniently applied to
 degenerate model spaces such as the $sd$ one-shell
one. \cite{dongzbox11}

 As mentioned earlier, the in-medium NNN potential $V_{3N}^{med}$ 
is dependent on the nuclear medium density $\rho$. We shall study
this $\rho$ dependence for  both low and moderately high densities. 
Clearly, at zero density $V_{3N}^{med}$ vanishes. 
Nuclear matter calculations
using $V_{2N}$ alone have not been able to satisfactorily reproduce
the empirical nuclear matter saturation properties \cite{siu09,dong09}.
It may be useful to study if the $\rho$ dependence of $V_{3N}^{med}$
 near the nuclear matter saturation density
$\rho_0$ (0.16 fm$^{-3}$) may play an important role 
for nuclear matter saturation. 
We shall do so by carrying out nuclear matter calculations 
with and without the inclusion of $V_{3N}^{med}$.
Similarly this dependence may also be important for closed-shell 
nuclei, such as $^{16}O$, whose nucleons
are embedded in a nuclear medium of densities near $\rho_0$. 
We shall calculate closed-shell
nucleus $^{16}O$ with and without the inclusion of  $V_{3N}^{med}$, 
as a further
 study of its effect near $\rho_0$. The valence nucleons, such as those
of $^{18}O$, are in a medium of densities much less than $\rho _0$. 
Thus our shell-model calculations for valence nuclei mentioned earlier
are a study of $V_{3N}^{med}$ at low densities.

The organization
of the present paper is as follows.
We shall describe some details about the derivation of $V_{3N}^{med}$
from $V_{3N}$ in section II. It may be noted that $V_{3N}^{med}$
is an effective density-dependent two-nucleon interaction, which
is more convenient than its underlying three-nucleon potential for 
nuclear many-body calculations. 
The methods we shall employ for the derivation of the shell-model
effective interactions $V_{eff}$ from $V_{2N}$ and $V_{3N}^{med}$
will  be outlined there. In section III we shall present first
our results of  nuclear matter calculations using $V_{2N}$ with and without
the inclusion of $V_{3N}^{med}$. The ring-diagram
formalism \cite{siu09,dong09} employed for our nuclear matter calculations
 will also be outlined. Next we shall
report the results of a similar ring-diagram calculation for $^{16}O$.
Results of our shell-model calculations of $^{18}O$, $^{18}F$,
 $^{19}O$ and $^{19}F$ using $V_{2N}$ with and without the inclusion
of $V_{3N}^{med}$ will  be presented and discussed in this section.
A summary and conclusion is presented in section IV.

\section{II. Formalism}

We first describe how we include the effects of the leading-order chiral 
three-nucleon interaction, $V_{3N}$, in our calculations. We consider the  
nuclear interaction $V$ as given by 
$V=(V_{2N}+V_{3N}^{\rm med})$, where $V_{2N}$ 
is  the N$^3$LO Idaho two-nucleon potential
\cite{idaho} and $V_{3N}^{\rm med}$ is a  
density-dependent two-body interaction
obtained from the chiral three-nucleon  force by closing 
one pair of external lines
and summing over the filled Fermi sea of nucleons. 
The leading contribution to
$V_{3N}$ occurs at N$^2$LO in the chiral power counting and is 
composed of a long-range two-pion exchange component $V_{3N}^{2\pi}$, 
a medium-range one-pion
exchange term $V_{3N}^{1\pi}$, and a pure contact interaction 
$V_{3N}^{ct}$:
\begin{equation}
V_{3N}^{(2\pi)} = \sum_{i\neq j\neq k} \frac{g_A^2}{8f_\pi^4} 
\frac{\vec{\sigma}_i \cdot \vec{q}_i \, \vec{\sigma}_j \cdot
\vec{q}_j}{(\vec{q_i}^2 + m_\pi^2)(\vec{q_j}^2+m_\pi^2)}
F_{ijk}^{\alpha \beta}\tau_i^\alpha \tau_j^\beta,
\label{3n1}
\end{equation}
\begin{equation}
V_{3N}^{(1\pi)} = -\sum_{i\neq j\neq k}
\frac{g_A c_D}{8f_\pi^4 \Lambda_\chi} \frac{\vec{\sigma}_j \cdot 
\vec{q}_j}{\vec{q_j}^2+m_\pi^2}
\vec{\sigma}_i \cdot
\vec{q}_j \, {\vec \tau}_i \cdot {\vec \tau}_j ,
\label{3n2}
\end{equation}
\begin{equation}
V_{3N}^{(\rm ct)} = \sum_{i\neq j\neq k} \frac{c_E}{2f_\pi^4 \Lambda_\chi}
{\vec \tau}_i \cdot {\vec \tau}_j,
\label{3n3}
\end{equation}
where $g_A=1.29, f_\pi = 92.4 {\rm MeV}, \Lambda_{\chi} = 700 {\rm MeV}, 
m_{\pi} = 138.04 {\rm MeV}/{\rm c}^2$ is the average pion mass, 
$\vec{q}_i=\vec{p_i}^\prime -\vec{p}_i$ is the difference between the 
final and initial momentum of nucleon \emph{i} and 
\begin{eqnarray}
\nonumber F_{ijk}^{\alpha \beta} &=& \delta^{\alpha \beta}\left (-4c_1m_\pi^2
 + 2c_3 \vec{q}_i \cdot \vec{q}_j \right )\\
&+& c_4 \epsilon^{\alpha \beta \gamma} \tau_k^\gamma \vec{\sigma}_k
\cdot \left ( \vec{q}_i \times \vec{q}_j \right ).
\label{3n4}
\end{eqnarray}
The low-energy constants $c_1 =-0.76 {\rm GeV}^{-1}$, 
$c_3=-4.78 {\rm GeV}^{-1}$, 
and $c_4 =3.96 {\rm GeV}^{-1}$ appear already 
in the N$^2$LO two-nucleon potential 
and are therefore constrained by low-energy NN phase shifts 
\cite{rentmeester}.
The low-energy constants $c_D$ and $c_E$ are typically fit 
to reproduce the
properties of light nuclei \cite{nogga00,pieper02,navratil07}.

A general three-body force may be written in second quantization as
\begin{equation}
\hat V_{3N}=\frac{1}{36}\sum_{123456}V([123],[456]) 
b^{\dagger}_1 b^{\dagger}_2 
b^{\dagger}_3 b_6 b_5 b_4, 
\end{equation}
where the antisymmetrized matrix element is
\begin{equation}
V([123],[456])\equiv
\langle 123|V_{3N}|456+645+564-654-546-465 \rangle.
\label{a3n}
\end{equation}
Here $\langle 123|V_{3N}|456 \rangle$ is a simple product matrix element,
and $b^{\dagger}$ and $b$ are creation and destruction operators defined 
with respect to the particle-hole vacuum $|C \rangle$ with $b_k |C\rangle$ 
= 0 for all \emph{k}. From eqs.\ (\ref{3n1}-\ref{3n4}) we can write 
$V_{3N} = V_{3N}^{(1)} + V_{3N}^{(2)} + V_{3N}^{(3)}$, where $V_{3N}^{(i)}$ 
is the component of $V_{3N}$ that is symmetric with respect to the interchange
$j \leftrightarrow k$. Now we contract one pair of the $b^{\dagger}$ and 
\emph{b} operators of the above $\hat V_{3N}$ (both operators must be holes), 
and this leads to an effective two-body force 
\begin{equation}
{\hat V}^{\rm med}_{3N}=\frac{1}{4}\sum_{1245}D([12],[45]) 
b^{\dagger}_1 b^{\dagger}_2  b_5 b_4, 
\end{equation}
with
\begin{eqnarray}
\nonumber D([12],[45])&=&\sum_{i\leq k_F}\left [ 
\langle i12|V^{(2)}_{3N}|i45 \rangle + 
\langle 1i2|V^{(2)}_{3N}|4i5 \rangle \right .\\
\nonumber &+& \langle 12i|V^{(2)}_{3N}|45i \rangle - 
\langle 1i2|V^{(2)}_{3N}|i45 \rangle \\
\nonumber &-& \langle i12|V^{(2)}_{3N}|4i5 \rangle - 
\langle 12i|V^{(2)}_{3N}|4i5 \rangle \\
\nonumber &-& \langle 1i2|V^{(2)}_{3N}|45i \rangle + 
\langle i12|V^{(2)}_{3N}|45i \rangle \\
&+& \left . \langle 12i|V^{(2)}_{3N}|i45 \rangle - 
(4 \leftrightarrow 5)\right ],
\end{eqnarray}
where $4\leftrightarrow 5$ denotes the nine exchange terms. 
The above  result  is unchanged when 
$V_{3N}^{(2)}$
is replaced by either $V_{3N}^{(1)}$
or $V_{3N}^{(3)}$. 
In our calculations we consider a background 
medium of symmetric nuclear matter at constant density 
characterized by a Fermi
momentum $k_F$. In this way analytic expressions can be obtained for 
$V_{3N}^{\rm med}$, as shown in refs. \cite{holt09,holt10}. The above is a 
density dependent effective `two-nucleon' interaction which, unlike its 
underlying three-nucleon force, can be readily used in many-body problems.
As detailed in \cite{holt10}, the partial-wave
potentials of the above $V_{3N}^{med}$ 
have been derived from the
lowest-order three-nucleon force. We shall use them in our
calculations, namely we shall consider the nucleon interaction as given by 
($V_{2N}+V_{3N}^{med}$).

We use the folded-diagram theory
\cite{ko90,jensen95,coraggio09,klr} to calculate the effective interaction.
 Briefly speaking, in this theory
$V_{eff}$ is given
 by a folded-diagram series 
\bq
V_{eff}= \hat{Q} - \hat{Q}^{'}\int\hat{Q}
+ \hat{Q}^{'}\int\hat{Q}\int\hat{Q} - \hat{Q}^{'}\int\hat{Q}\int\hat{Q}
\int\hat{Q}\cdots,
\eq
where $\hat Q$ represents a so-called $\hat Q$-box consisted of
irreducible diagrams, as  illustrated by the 1st- and 2nd-order
diagrams of Fig. 1. 
 (The $\hat Q'$-box is the same as the $\hat Q$-box except that
 $\hat Q'$ does not have diagrams 1st-order in the interaction.) 
Suppose we include only $V_{2N}$. Then each vertex in the diagrams 
shown represents
a $V_{low-k}$ low-momentum interaction 
\cite{bogner01,bogner02,bogner03} derived from $V_{2N}$. 
When we include also $V_{3N}^{med}$ in the calculation of $V_{eff}$, 
the $\hat Q$-box diagrams for
effective interactions will have a specific type of $V_{3N}^{med}$ vertices
as illustrated in Fig. 2.
 Recall that diagram d4 of Fig. 1 represents the interaction between 
two valence nucleons via $V_{2N}$. This diagram becomes
diagram (a) of Fig. 2 if its $V_{2N}$ is replaced by $V_{3N}^{med}$. 
Here the two valence nucleons have to have the participation of a sea-nucleon 
(below Fermi sea) in order to
activate $V_{3N}$, as indicated by the hole-line loop in the diagram.
( Note each $V_{3N}^{med}$ vertex of Fig. 2 represents a $V_{low-k}$ 
interaction derived from $V_{3N}^{med}$.)
Similarly diagram d1 of Fig. 1 represents
the interaction of a valence nucleon with a sea-nucleon 
 via the two-nucleon interaction $V_{2N}$. If this interaction
is replaced by $V_{3N}^{med}$, this diagram becomes diagram (b) of Fig. 2.
$V_{3N}$ must involve three nucleons, and hence here the valence nucleon
interacts with `two' sea-nucleons as indicated by the two hole-line
loops attached to $V_{3N}$. 
Diagram (c) is the $V_{3N}$ core polarization diagram 
corresponding to diagram
d7 of Fig. 1. Again  here one hole-line loop is needed for each vertex 
to activate $V_{3N}$. 
 As seen from Eqs. (31) and (32),  $V_{3N}^{med}$ requires the 
involvement of at least one
sea-nucleon, and it is this requirement which
is reflected by  the hole-line loops
attached to the $V_{3N}$ vertices in Fig. 2. To summarize,  the 
$\hat Q$-box diagrams of Fig. 1 are used for calculating $V_{eff}$
when we use only $V_{2N}$, while additional diagrams as illustrated
by Fig. 2 are also included when the interaction is consisted of
both $V_{2N}$ and $V_{3N}$.

\begin{figure}
\scalebox{0.3}{
\includegraphics[angle=-90]{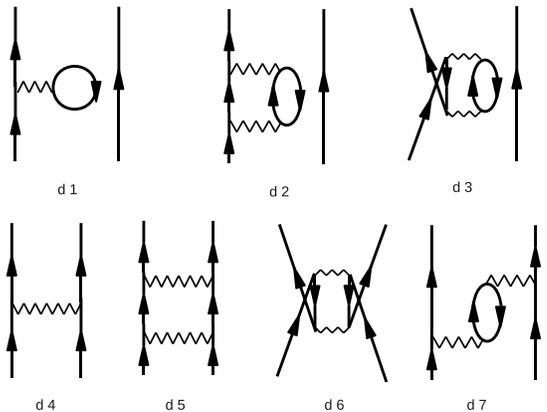}
}
\caption{Low-order diagrams constituting the $\hat Q$-box.}
\end{figure}

\begin{figure}
\scalebox{0.33}{
\includegraphics[angle=-90]{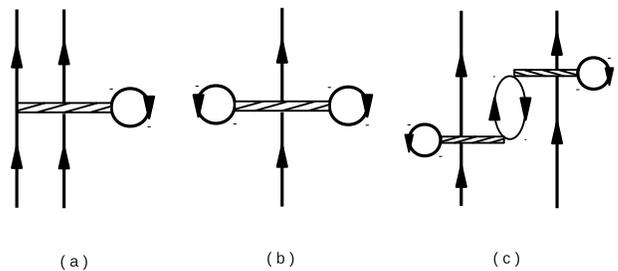}
}
\caption{$\hat Q$-box diagrams with three-nucleon forces.  Each $V_{3N}^{med}$ 
interaction
is represented by a `narrow band with hole-line-loop' vertex.}
\end{figure}

 After obtaining the $\hat Q$-box, we calcualte the effective interaction
of Eq. (9) using iteration methods. The {\rm LS} method \cite{lesu,sule} is 
convenient for degenerate model spaces, and it is used to calculate
the \emph{sd} one-shell effective interactions. For the non-degenerate
\emph{sdpf} two-shell effective interactions, one can use either
 the Krenciglowa-Kuo iteration method ({\rm KK}) \cite{kren,kuo95}
or the recent extended {\rm KK} method ({\rm EKKO}) of Okamoto {\it et al.} 
\cite{okamoto10}. Both the EKKO and KK methods are convenient
for calculating the effective interaction for non-degenerate
model spaces, while the {\rm EKKO} method being more efficient
(faster converging rate). \cite{dongzbox11}.
The main difference between the {\rm EKKO} and {\rm KK} methods 
is the following.
The {\rm KK} method employs the irreducible vertex function $\hat Q$-box
which may be written as
\bq
\hat Q(\omega)=[PVP+PVQ\frac{1}{\omega -QHQ}QVP]_L
\eq
where {\rm V} represents the {\rm NN} interaction, {\rm P} 
denotes the model-space
projection operator, and $Q \equiv (1-P)$. The energy variable $\omega$ is 
 determined self-consistently \cite{lesu,sule}. 
Note that $\hat Q$ contains only
valence-linked diagrams as indicated by the subscript {\it L}.
 In the EKKO method, the above $\hat Q$-box is
replaced by the $\hat Z$-box defined by \cite{okamoto10,dongzbox11}
\begin{equation}
\hat Z(\omega)=\frac{1}{1-\hat Q_1(\omega)}[\hat Q(\omega)
-\hat Q_1 (\omega)P(\omega-H_0)P],
\end{equation}
where $\hat Q_1$ is the first derivative $d\hat Q/d \omega$. When the P-
and Q-space are not sufficiently separated, the above $\hat Q$-box
may have singularities. An important
advantage of the {\rm EKKO} method is that the $\hat Z$-box is well behaved
even when the corresponding $\hat Q$-box is singular, overcoming
the above singularity difficulty and making the {\rm EKKO}
method more efficient 
\cite{okamoto10,dongzbox11}. We shall calculate the \emph{sdpf} two-shell
effective interactions using both the {\rm EKKO} and {\rm KK} 
iteration methods, to cross check their convergences.

\section{III. Results and discussion}

As described in section II, we shall calculate
the shell-model effective interactions with the inclusion of
the medium-dependent three-nucleon
force $V_{3N}^{med}$ which is obtained from a chiral {\rm NNN} force 
by integrating one participating sea-nucleon over the Fermi sea.
The $V_{3N}^{med}$ interaction is a density dependent 
effective interaction, and 
we shall first carry out nuclear matter calculations
with the inclusion of $V_{3N}^{med}$, to study the properties of 
this interaction at various
densities near the nuclear matter saturation density $\rho_0$.
Before doing so, we need to choose or decide the low-energy constant 
of $V_{3N}$ (see Eqs. (1-3)) to be used in our calculations.
The low-energy constants
 $c_1$, $c_3$ and $c_4$ of $V_{3N}^{(2 \pi)}$
are well determined; they are constrained by low-energy 
{\rm NN} scattering data \cite{rentmeester}.
Their values so determined (section II) will be used in the present work.
But the low-energy constants $c_D$ and $c_E$, of $V_{3N}^{(1\pi)}$
and $V_{3N}^{(ct)}$ respectively, are less well known; their values
 determined from properties of light nuclei
exhibit considerable
variations \cite{nogga00,pieper02,navratil07,holt09}.
The $V_{3N}^{med}$ interaction  depends
explicitly on the Fermi momentum $k_F$ which is well defined for
nuclear matter. But $k_F$ is not well defined  for finite nuclei,
causing uncertainty in determining the low-energy constants
of $V_{3N}^{med}$ from properties of finite nuclei.
But for nuclear matter, $k_F$ is well defined; we have thus chosen 
to study these constants
by way of  nuclear matter calculations with  $V_{3N}^{med}$ included.


\begin{figure}[hbt]
\scalebox{0.3}{
\includegraphics[angle=-90]{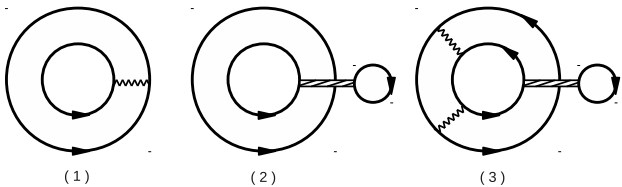}
}
\caption{Nuclear matter ring diagrams with vertices from $V_{2N}$
(wavy line) and $V_{3N}^{med}$ (narrow band  with hole-line-loop). }
\end{figure}
            
We calculate the equation of state ({\rm EOS}) 
for symmetric nuclear matter using
a ring-digram formalism \cite{siu09,dong09}. A brief description
of this formalism is presented below, to outline how we 
include $V_{3N}^{med}$ in our calculations.
Using familiar renormlization procedures \cite{bogner01,bogner02,bogner03},
we first calculate  the low-momentum
interactions $V_{low-k}^{2N}(\Lambda)$ and $V_{low-k}^{3N}(\Lambda)$ 
respectively from $V_{2N}$ \cite{idaho} and $V_{3N}^{med}$.
A common decimation scale 
of $\Lambda = 2.1 {\rm fm}^{-1}$ is employed. 
This value is chosen because
at this scale the low-momentum interactions derived from
different {\rm NN} potentials \cite{idaho,cdbonn,argonne,nijmegen}
are remarkably close to each other \cite{bogner03}, leading to a
 nearly unique low-momentum interacton. 
 Using the above ring-diagram framework, the ground-state 
energy shift $\Delta E$ 
is given by the all-order sum of the \emph{pphh} ring diagrams 
as illustrated
in Fig. 3. ($\Delta E_0$ is defined as ($E_0-E_0^{free}$) where $E_0$
is the  true ground state energy and $E_0^{free}$ that for the non-interacting
system.) As shown in Fig.3, diagram (1) is a 1st-order  ring diagram  
with its vertex $V_{2N}^{low-k}$ calculated from the N$^3$LO chiral NN
potential \cite{idaho}. Similarly diagram (2) is a 1st-order ring diagram
with $V_{3N}^{low-k}$ obtained from 
the lowest order chiral $V_{3N}$ as described  in section II. Diagram (3)
has vertices from both $V_{2N}^{low-k}$ and $V_{3N}^{low-k}$.
Similar to Fig. 2, to have the pair of nucleons in the ring diagrams
interact with $V_{3N}$ there must be the participation of a third nucleon.
Thus the $V_{3N}$ vertices in Fig. 3 all have a one-hole-line loop
attached to them.

With these ring diagram summed to all orders, the ground-state energy 
shift for nuclear matter is given as 
\begin{eqnarray}\label{eng}
\Delta E_0&=&\int_0^1 d\lambda
\sum_m \sum_{ijkl<\Lambda}Y_m(ij,\lambda) Y_m^*(kl,\lambda)\nonumber \\
&& \times  \langle
ij|[V_{low-k}^{2N}(\Lambda)+V_{low-k}^{3N}(\Lambda)]|kl \rangle.
\end{eqnarray}
The transition amplitudes above are 
$Y_m^*(kl,\lambda)=\langle \Psi_m(\lambda,A-2)|\beta (kl)
|\Psi_0(\lambda,A)\rangle$,
where $\Psi_0(\lambda,A)$ denotes the true ground state of nuclear matter
which has \textit{A} particles, 
 $\Psi_m(\lambda,A-2)$ the \emph{m}th 
true eigenstate of the (\textit{A}-2) system,
and $\beta (kl)=b_lb_k$ if $(k,l)>k_F$ and $=b_k^{\dagger}b_l^{\dagger}$
if $(k,l)\leq k_F$.
Note that $\lambda$ is a strength parameter, integrated from
0 to 1.  The  amplitudes \emph{Y}
 are calculated from a {\rm RPA} equation \cite{siu09,dong09,song87} based on
the ($V_{2N}$+$V_{3N}^{med}$) interactions. 
 Note that our calculation is the same
as the usual Hartree-Fock ({\rm HF}) one if we include only the 
1st-order ring diagrams (1) and (2)
 of Fig. 4, and in this case the above
\emph{Y} amplitude becomes $Y(ij)=n_in_j$ where $n_i$=1 for $i<k_F$ and =0
otherwise. Our ring-diagram calculations include particle-hole
fluctuations of the Fermi sea, while not so for the {\rm HF} calculations.

\begin{figure}[hbt]
\scalebox{0.42}{
\includegraphics[angle=-90]{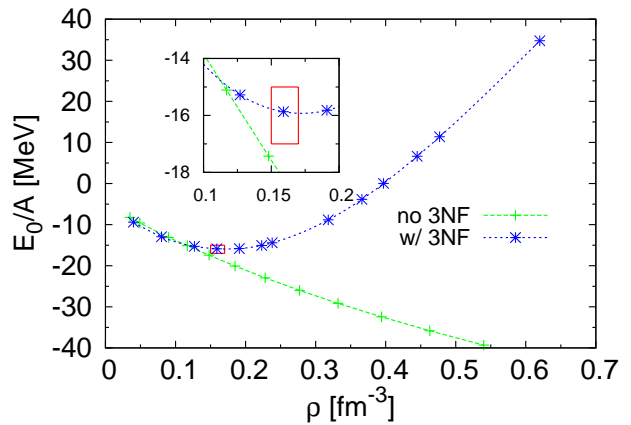}
}
\caption{Ring-diagram equation of state for symmetric nuclear matter 
calculated with (w/ 3NF) and without (no 3NF) $V_{3N}^{med}$.}
\end{figure}

We have carried out ring-diagram calculations for symmetric nuclear
matter using a wide range of values for $c_D$ and $c_E$.
Our results with  
$c_D=-2.7$ and $c_E=0.7$ are displayed in Fig. 4; they give 
$E_0/A \simeq -16 {\rm MeV}$
and $\rho_0\simeq 0.16 {\rm fm}^{-3}$, both in satisfactory 
agreement with the empirical values. Two curves are 
shown in the figure, one with $V_{2N}$ alone and the other
with the addition of $V_{3N}^{med}$. Comparing with the $V_{2N}$ curve,
it is of interest that the effect of $V_{3N}^{med}$ is slightly
attractive for low densities ($< \sim 2\rho_0/3$)
while becomes strongly repulsive at high densities. Recall that we have used
a common decimation scale of $\Lambda=2.1 {\rm fm}^{-1}$ \cite{bogner03}  
for both $V_{2N}^{low-k}$ and
$V_{3N}^{low-k}$. It is of interest that in the ring-diagram calculations
with Brown-Rho scaling \cite{siu09,dong09}  a large
$\Lambda$ of $\sim 3.0 {\rm fm}^{-1}$ is needed for obtaining satisfactory
nuclear matter saturation properties, while in the present calculation
with $V_{3N}^{med}$ satisfactory results can be obtained with the  use of
 a smaller $\Lambda$ (2.1 ${\rm fm}^{-1}$).


 We have calculated the {\rm HF} s.p. spectrum $\epsilon _k$ in nuclear matter
with the inclusion of both $V_{2N}$ and three-nucleon force $V_{3N}^{med}$.
The inclusion of the latter has been found to have significant
effect to the spectrum.  $\epsilon _k$ can be well fitted by the quadratic
expression ($k^2 \hbar^2/(2m^*)+\Delta$) where $m^*$ is the effective mass
and $\Delta$ is a well-depth parameter representing the s.p. energy
at zero momentum. In Fig. 5 we present our
results for $\Delta$ and $m/m^*$ for various densities. Two curves are
shown: the lower one without and  the top one with the inclusion of
$V_{3N}^{med}$. The densities for the 7 data points of each curve
are, from left to right, (0.25, 0.5, 1.0, 1.5, 2.0, 2.5, 3.0 $\rho_0$) 
respectively. 
As shown by the lower curve of  the figure, 
$\Delta$ and $m/m^*$ both vary monotonically
with the density when only $V_{2N}$ is employed. With increasing density,
$\Delta$ becomes incresingly more negative and $m/m^*$ increasingly
larger, exhibiting no saturation (up to 3.0 $\rho_0$). 
The trend is quite different for the 
upper curve where the three-nucleon force is included. 
Here $m/m^*$ still monotonically increases with density, but
$\Delta$ arises after saturation,  approaching zero at
  some higher density. It is of interest that  $V_{3N}^{med}$ has a large 
effect in raising the chemical potential of nucleons
 in high density nuclear matter; this raise of the chemical potential 
may play an important role in enabling such nucleons  
 decaying into other baryons such as hyperons.

\begin{figure}[here]
\scalebox{0.42}{
\includegraphics[angle=-90]{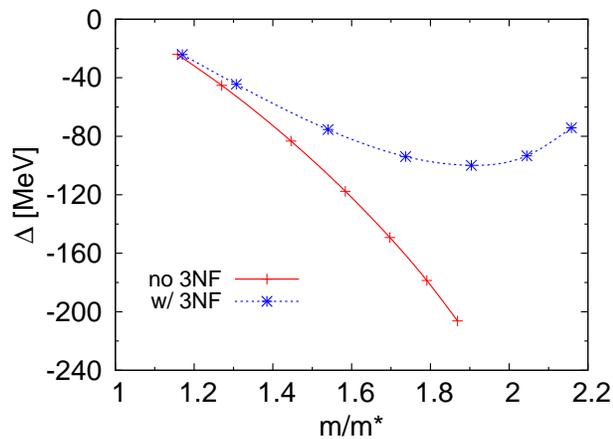}
}
\caption{Comparison of $m^*$ and $\Delta$ for the in-medium
s.p. spectrum calculated with (w/ 3NF) and without (no 3NF) $V_{3N}^{med}$.
See text for other explanations.}
\end{figure}

  As indicated by  Fig. 4, our ring-diagram calculations with the 
inclusion of the {\rm NNN}
force $V_{3N}^{med}$ have described well the nuclear matter saturation
properties. It would be useful and of interest to check if this approach 
is also satisfactory for describing closed-shell nuclei whose nucleons
 are mainly embedded in a similar nuclear medium of densities near $\rho _0$,
the nuclear matter satturation density. 
To study this possiblity, we have extended the above
 ring-diagram nuclear matter calculation to the closed shell nucleus
$^{16}{\rm O}$, using the same low-energy parameters 
$C_D$ and $C_E$ (-2.70 and 0.70)
determined earlier. Note that these parametrs will also be used later
in our $sd$ and $sdpf$ shell-model calculations.  
Our ring-diagram calculation for $^{16}O$
is quite similar to the nuclear matter
one. The ground-state energy shift for $^{16}{\rm O}$ is also given by
the all-order sum of the ring diagrams illustrated in Fig. 3, 
except for the following
differences.  For nuclear matter, the s.p. states are plane-wave states and
each particle line in the figure denotes a plane-wave
particle with momentum $k>k_F$ and each hole line with $k<k_F$. 
For $^{16}{\rm O}$
we use harmonic oscillator s.p. wave functions ($\hbar \omega$=14 {\rm MeV}),
and each particle line in the figure denotes a nucleon in the 
particle shells (\emph{1s0d1p0f}), 
and each hole line a nucleon in the \emph{0s0p} shells. 
For nuclear matter, we use  momemtum model space where all particles
have momentum $k<\Lambda$. For $^{16}{\rm O}$, we use an oscillator model
space  composed of the s.p. orbits of the $0s0p1s0d1p0f$ shells.
Indeed the ring-diagram calculations for nuclear matter and $^{16}{\rm O}$
are quite similar. The ground-state energy shift for $^{16}{\rm O}$ is also
given  by Eq.(12), except that the orbits (\emph{ijkl}) now all refer to
shell-model orbits  and the summation restriction of
 $(ijkl<\Lambda)$ is replaced by that of \emph{ijkl} belonging 
to the above oscillator model space.

 With the same $V_{2N}$ and $V_{3N}^{med}$ interactions used for our nuclear
matter calculations, we have carried out ring-diagram calculations
for the ground-state energy of $^{16}{\rm O}$. 
Since $V_{3N}^{med}$ is dependent on the nuclear density $\rho$,
 we need to have the value of  $\rho$ to carry on the calculation.
For nuclear matter, this $\rho$ is well defined, but not so for finite
nuclei. A preliminary estimate for the density for $^{16} {\rm O}$ can be made
by considering it as a uniform sphere \cite{ringschuck};
in this way the average density for $^{16}O$ is estimated as 
 $\sim 0.85 \rho_0$. A more realistic estimate for this density 
should be lower, as physically
the nuclear surface  is diffused (not sharp).
In Table I we present results for the ground-state energy per 
nucleon of $^{16}{\rm O}$ for sevearl densities,  
$\rho/\rho_0=$ 0, 0.5, 0.70 and 0.75. 
As seen, the binding energy calculated with $\rho =0$ (i.e. no $V_{3N}^{med}$)
is clearly not adequate. The inclusion of $V_{3N}^{med}$ lowers the 
ground-state energy per nucleon significantly; for example
 the inclusion of $V_{3N}^{med}$ 
at $\rho/\rho_0$=0.75 has increased the ring-diagram ground-state energy
 from -6.97 (only $V_{2N}$) to -8.01 MeV (with $V_{3N}^{med}$) per nucleon,
 in good agreement with experiment.
 Since we have not included the
Coulomb interaction in our calculations,  an empirical
Coulomb energy of 1.14 {\rm MeV} per nucleon \cite{ringschuck} 
has been included
in the above calculation for the ground-state energy of $^{16}{\rm O}$.
It may be noted that our results for $\rho/\rho_0$= 0.70 and 0.75
are rather close to each other. 
We have calculated the potential
enengy({\rm PE}) from the $(V_{2N}+V_{3N}^{med})$ interaction. 
Results given by the
1st- and 2nd- and all-order ring diagrams, denoted by PE-1st, PE-2nd
and PE-ring respectively, are listed in the Table.
It may be noted that PE-2nd is much smaller than PE-1st. And the difference
between (PE-1st+PE-2nd) and PE-ring is rather small, about 5$\%$.
Bogner {\it et al.} \cite{bogner05} have found that $V_{low-k}$ is suitable
for perturbative nuclear matter calculations. The good convergence
shown in Table I suggests that this interaction is also suitable for
perturbative calculations of finite nuclei.

\begin{table}
\caption{Ring-diagram calculations of the ground-state energy per nucleon
$E_0$ for $^{16}{\rm O}$. The interaction ($V_{2N}+V_{3N}^{med}$) is used.
The potential energy per nucleon given by the 1st-, 2nd- and all-order
ring diagrams are denoted respectively by PE-1st, PE-2nd and
PE-ring. $E_0-1st$ and $E_0-ring$ are given by the 1st- and all-order
ring diagrams respectively. The experimental result is from 
\cite{nucldata}. All energies are in {\rm MeV}.}
\begin{center}
\begin{tabular}{|c|c|c|c|c|c|}\hline
     $\rho/\rho_0$ & PE-1st &PE-2nd& PE-ring &$E_0$-1st&$E_0$-ring\\ 
    \hline
     0&     -19.84 &-1.99 &-23.02&-3.61&-6.79  \\
     0.50&   -20.67 & -2.28  &-23.60 &-4.44&-7.37 \\
     0.70&   -20.78 & -2.38  &-24.11 &-4.55&-7.87 \\
    0.75 &  -20.79 & -2.40 & -24.25&-4.56&-8.01   \\ \hline
    $E_o$-exp&&&&& -8.00\\ \hline  
\end{tabular}
\end{center}
\end{table}



 Having seen that the medium-dependent three-nucleon force
 $V_{3N}^{med}$ has given desirable results for nuclear matter
and closed-shell nucleus $^{16}O$, we shall now study its effects
on the nuclear effective interactions
for the degenerate  \emph{sd} one-shell and the non-degenerate 
 \emph{sdpf} two-shell model spaces. We use the same methods as described
in \cite{dongzbox11} for the calculation and application of
 these effective interactions, except
for the following difference: In \cite{dongzbox11} only the two-nucleon
interaction $V_{2N}$ was employed. In the present work we include
both $V_{2N}$ and $V_{3N}^{med}$.
We first calculate the low-momentum interactions from these
interactions, and using them to obtain the $\hat Q$-box diagrams
as shown in Figs. 1 and 2. 
The same $V_{2N}$ and $V_{3N}^{med}$
interactions we used earlier in our calculations for nuclear matter
and $^{16}O$ are employed.
The degenerate $sd$  and non-degenerate $sdpf$  effective interactions are 
obtained from the $\hat Q$-box
using respectively the LS iteration method \cite{lesu,sule} and the 
EKKO iteration method \cite{okamoto10,dongzbox11}.
The nondegenerate \emph{sdpf} effective interactions can also be calculated
using the KK iteration method \cite{kren,kuo95,dongzbox11}.

\begin{figure}
\scalebox{0.42}{
\includegraphics[angle=-90]{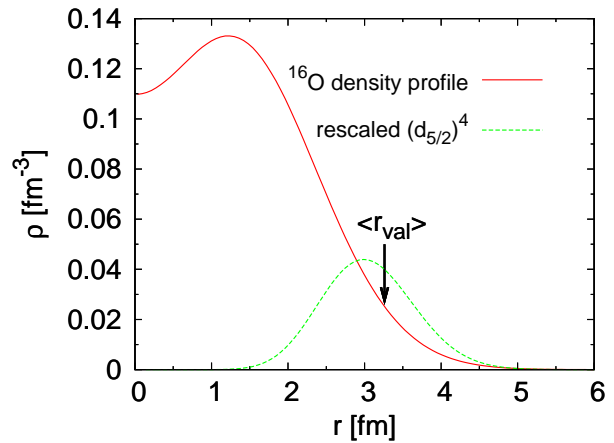}
}
\caption{ Density profile of $^{16}{\rm O}$ (solid line) and the radial
distribution of the fourth power of the \emph{0d} shell-model wave function
(dashed line). See text for other explanations. }
\end{figure}

\begin{figure}[here]
\scalebox{0.42}{
\includegraphics[angle=-90]{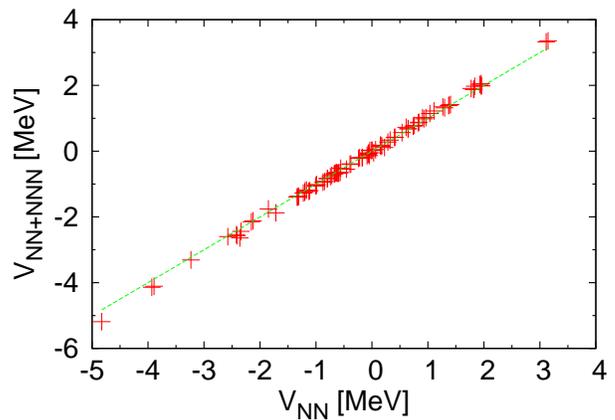}
}
\caption{ Comparison of the $sd$-shell matrix elements of $V_{eff}$
calculated from $V_{2N}$ alone (NN) and from 
$V_{2N}+V_{3N}$ (NN+NNN). }
\end{figure}

\begin{figure}
\scalebox{0.42}{
\includegraphics[angle=-90]{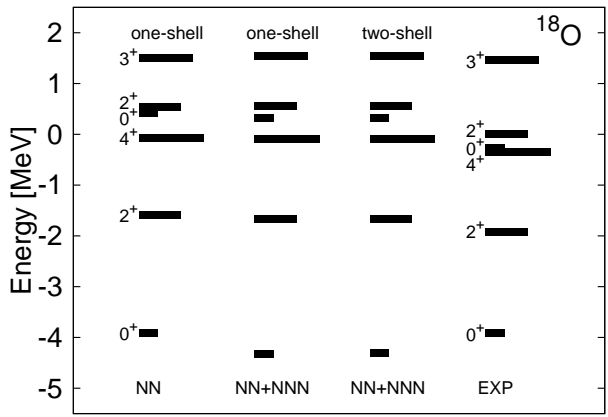}
}
\caption{ Energy spectrum of $^{18}O$ calculated from
$V_{2N}$ alone (NN) and from $V_{2N}+V_{3N}$ (NN+NNN). The $sd$ (one-shell)
and $sdpf$ (two-shell) model spaces are both employed. 
The experimental results are from \cite{nucldata}.  }
\end{figure}

\begin{figure}
\scalebox{0.42}{
\includegraphics[angle=-90]{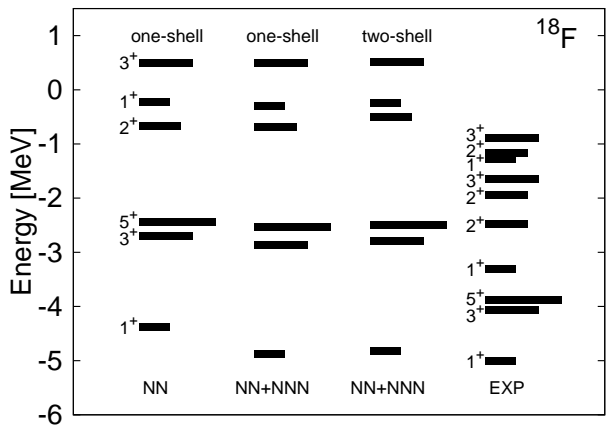}
}
\caption{Same as Fig. 8 except for $^{18}{\rm F}$. }
\end{figure}

 We first apply the above effective interactions with $V_{3N}^{med}$ 
to nuclei 
$^{18}O$ and $^{18}F$. Since $V_{3N}^{med}$ is density dependent, 
 we  need to know the `local density' $\rho _v$ felt by
 the  valence nucleons in these nuclei. It should be small, but its precise
value is rather uncertain. In the present work, we shall estimate $\rho _v$
by comparing the density profile of $^{16}O$ and that for the valence
nucleons. In Fig. 6 we plot the density profile 
$\rho_{core}(r)$ of $^{16}O$
obtained with its wave function assumed to be a closed 
shell-model $s^4p^{12}$ 
core. We  also plot the  distribution $\Phi(r)=\xi \phi(r)^4$ where
$\phi(r)$ is the radial harmonic oscillator wave function for $d_{5/2}$,
with the scaling parameter $\xi$=0.1. 
As seen from the
plots, the pair of $d_{5/2}$ nucleons reside primarily in the low-density
region of $\rho_{core}$. Depending on the  averaging procedure employed,
we have estimated $\rho _v$ from  
$\rho _{core}(r)$ and $\phi (r)$, obtaining values ranging from $\sim 0.015$ 
to $\sim 0.030 fm^{-3}$. 
We have considered another scheme to estimate $\rho _v$. The rms radius
$<r_{val}>$ for the shell-model $d_{5/2}$ orbit is indicated by an arrow
in the figure. A simple scheme to estimate $\rho _v$ is to let
it equal to $\rho_{core}(<r_{val}>)$, the value obtained in this way
being $\sim 0.025 fm^{-3}$. This value may be reduced if realistic
s.p. wave functions are employed. 
The $d_{3/2}$ orbit is nearly unbound, and its rms radius  should be
 considerably larger than that for $d_{5/2}$. Assuming $\hbar \omega$
=10 MeV for the $d_{3/2}$ orbit, its $<r_{val}>$ would be $\sim 3.8 fm$
 giving $\rho _v \simeq 0.01 fm^{-3}$ for this orbit. 
We believe that a suitable range
for $\rho _v$ is from $\sim 0.1$ to $\sim 0.2 \rho_0$ ($\rho_0 \equiv
0.16 fm^{-3}$). In the present work, we adopt $\rho _v=0.15 \rho _0$,
to illustrae the effects of $V_{3N}^{med}$. 

 In Fig. 7  the matrix elements of the $sd$-shell effective interactions
calculated from $V_{2N}$ with and without $V_{3N}^{med}$, 
denoted respectively as 'NN+NNN' and 'NN', are compared.
As seen, the two sets of matrix elements are rather similar to each
other, with the magnitudes of those including $V_{3N}^{med}$ being slightly 
larger. The low-lying spectrum of $^{18}O$ and $^{18}F$ calculated with
the above interactions are presented in Figs. 8 and 9.
The main effect of the `NNN' force is a small downward shift for the
ground states of $^{18}O$ and $^{18}F$, while leaving the other states
largely unchanged. This trend is consistent with what we have observed
for nuclear matter calculations (see Fig. 4), namely the effect of
$V_{3N}^{med}$ is slightly attractive at low densities. 
With the inclusion of thre 'NNN' force, the ground-state energy
of $^{18}F$ is in good agreement with experiment while that for
$^{18}O$ is slightly overbound. The agreements of the other states
with experiments for $^{18}O$ are better than $^{18}F$.

  Using the method outlined in section II and  \cite{dongzbox11}, we have 
 also calculated the effective interaction for a  $sdpf$ 
two-shell space  using  $V_{2N}$ with and without 
the inclusion of  $V_{3N}^{med}$. In Figs. 8 and 9, the spectra 
of $^{18}O$ and $^{18}F$ obtained
with the former effective interaction are also presented 
('two-shell', 'NN+NNN'). 
As seen, the 'two-shell' and 'one-shell' results are rather similar
to each other. To further study the effects of the three-nucleon force 
$V_{3N}^{med}$, we have extended the calculations of Figs. 8 and 9
to nuclei $^{19}O$ and $^{19}F$, with results presented in Figs. 10 and 11.
Here, similar to $^{18}O$ and $^{18}F$, the ground-state energies of
$^{19}O$ and $^{19}F$ obtained with $V_{2N}$ only are both too high 
compared with experiments (see the 'NN' columns). As shown by the
'NN+NNN' columns, the inclusion of $V_{3N}^{med}$ slightly lowers
these energies. The 'one-shell' and 'two-shell' results are rather close
to each other for  
$^{19}O$.  But they are slightly different for  $^{19}F$, with the
'NN+NNN' ground-state energy noticebly lower than the 'NN' one.

 As shown in Figs. 8-11, although there are  qualitative agreements 
 between the
calculated and experimental spectra of $^{18}O$, $^{18}F$, $^{19}O$
and $^{19}F$, there are significant disagreements between them.
For example, the orderings of the calculated low-lying states of $^{19}O$
are in fair agreement with the experimental ones, 
but their relative spacings
are not. As shown by the figures, the effects 
from $V_{3N}^{med}$ are generally small and attractive, mainly slightly
lowering their ground-state energies while having nearly no influence
on the relative spacings of the calculated spectrum. 
It should be useful to further study how to reduce the above disagreements.
 We have employed
a low-order approximation for the $\hat Q$-box, including
only 1st- and 2nd-order diagrams. The inclusion of higher-order diagrams
may alter our results. As carried out by 
by Holt et al. \cite{jdholt05}, certain classes of planar diagrams of the
$\hat Q$-box can be summed up to all orders using the Kirson-Babu-Brown
induced interaction method. We plan to extend our present calculations
by including these planar diagrams, and study
their effects to the shell-model description of these nuclei.
?

\begin{figure}
\scalebox{0.42}{
\includegraphics[angle=-90]{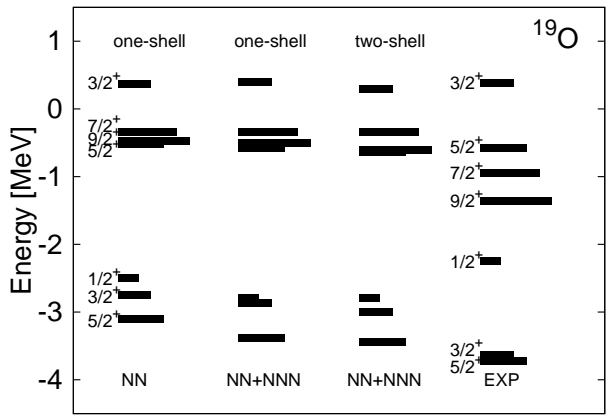}
}
\caption{ Same as Fig. 8 except for  $^{19} O$. }
\end{figure}

\begin{figure}
\scalebox{0.42}{
\includegraphics[angle=-90]{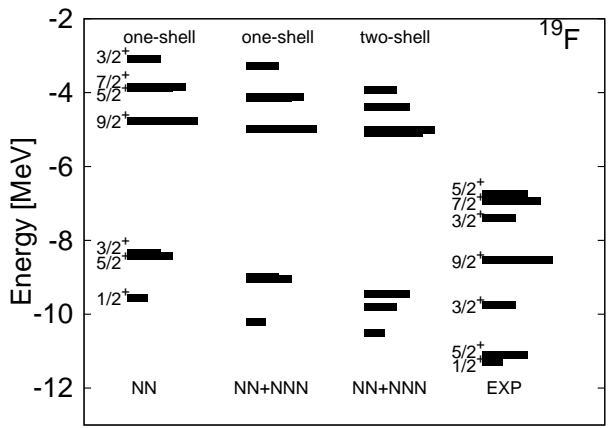}
}
\caption{Same as Fig. 8 except for $^{19}F$.}
\end{figure}

\section{IV. Summary and conclusion}
 We have carried out shell-model calculations for nuclei $^{18}O$, $^{18}F$,
$^{19}O$, $^{19}F$ and $^{16}O$ using effective interactions derived from
the chiral two- and three-nucleon potentials.
We start from the 
$N^3LO$ two-nucleon potential $V_{2N}$ 
together with an in-medium three-nucleon potential $V_{3N}^{\rm med})$ which
is a density-dependent effective interaction derived
from the lowest-order chiral three-nucleon force $V_{3N}$ 
by Holt, Kaiser and Weise \cite{holt09,holt10}.
The $V_{low-k}$ low-momentum interactions are then derived from these
potentials, and are used in a folded-diagram expansion for calculating
the $V_{eff}$ interactions for both the degenerate $sd$ and 
non-degenerate $sdpf$ model
soaces. The well-known Lee-Suzuki iteration
method \cite{lesu,sule} is used for the derivation of the $sd$ effective
interactions, while the recently developed iteration method
of Okamoto {\it et al.} \cite{okamoto10} is employed for the  $sdpf$ ones.

We have studied  the effects of  $V_{3N}^{med}$, which is an
effective two-body interaction depending on the medium density $\rho$,
at various densities  near and below the nuclear matter saturation 
density $\rho _0$.
We first apply the $V_{3N}^{med}$ interaction to nuclear matter 
calculations for which the medium density is uniform and
 well defined. A ring-diagram method for nuclear matter
has been employed. The low-energy constants $C_D$ and $C_E$, 
of respectively the one-pion and contact terms of the leading-order 
chiral three-nucleon
potential $V_{3N}$, are not well known. We have chosen these constants
by requiring that  their use in our nuclear matter calculations
with $V_{2N}$ and $V_{3N}^{med}$, which is derived from $V_{3N}$,
 satisfactorily reproduces the saturation 
properties of nuclear matter.
As indicated in Fig. 4,
the inclusion of $V_{3N}^{med}$ has largely improved our nuclear matter results
comparing with those given by $V_{2N}$ alone, and  the resulting nuclear matter
saturation propertiesa are in good agreement with the empirical values.

We then  calculate the $sd$ and $sdpf$ effective interactions with the 
inclusion of the above $V_{3N}^{med}$,  and apply
them to nuclei $^{18}O$, $^{18}F$, $^{19}O$ and $^{19}F$. An ambiguity
here is the determination of the local density $\rho _v$ felt by the
valence nucleons of these nuclei. This density should be small compared with
the saturation density $\rho _0$ of nuclear matter, but it is difficult 
to determine its value precisely. We have estimated this value based on the
density profile of $^{16}O$ given by a simple $s^4p^{12}$ shell-model
wave function. This is an approximation which should be further investigated
and improved upon.
In our calculations for these nuclei, we have adopted the 
estimation of $\rho_v/\rho_0$=0.15.
As indicated by Figs. 8-11, the main effect of $V_{3N}^{med}$ is to 
lower the ground-state energies
of these nuclei slightly, making them in better  agreements with experiments
than the energies given by $V_{2N}$ alone.
 The effects of $V_{3N}^{med}$ to these nuclei  are general small,
mainly because their valence nucleons are in a low-density medium.
 
The above $V_{3N}^{med}$ has also been applied to a shell-model calculation
of closed-shell nucleus $^{16}O$. The nucleons in closed-shell nuclei
 are embedded in nuclear medium of much larger density 
 than that for the valence nucleons  mentioned above.
For them the effect of $V_{3N}^{med}$ is expected be more important.
Indeed this is confirmed by our calculations.  
As illustrated in Table I,  the inclusion of 
this interaction has been
essential for satisfactorily  reproducing the 
empirical binding energy  of $^{16}O$.
It should be useful and of interest to  further study  
this in-medium three-nucleon
potential by applying it to closed-shell nuclei and those with a few
valence holes.

\begin{acknowledgments}
We are very grateful to L. Coraggio, A. Covello, 
 A. Gargano, N. Itaco, M. Machleidt and R. Okamoto  for many 
helpful discussions. 
Partial supports from the US Department of Energy under contracts
DE-FG02-88ER40388 and the DFG (Deutsche Forschungsgemeinschaft)
 cluster of excellence: Origin and Structure of the Universe are
 gratefully acknowledged.
\end{acknowledgments}


\begin{thebibliography}{99}
\bibitem{brownwild88} B.A. Brown and B.H. Wildenthal, Ann. Rev. Nucl. 
Part. Sci. {\bf 38}, 29(1988).
\bibitem{brownrich06} B.A. Brown, W.A. Richter, Phys. Rev. C74, 0343150 (2006).
\bibitem{jensen95} M.\ Hjorth-Jensen, T.\ T.\ S.\ Kuo, and E.\ Osnes,
 Phys.\ Rep.\ {\bf 261}, 126 (1995), and references therein.
\bibitem{coraggio09} L. Coraggio, A. Covello, A. Gargano, N. Itaco and
T.T.S. Kuo, Prog. Part. Nucl. Phys., 62 (2009) 135, and references
quoted therein.
\bibitem{siu09}L.W. Siu, J.W. Holt, T.T.S. Kuo and G.E. Brown, Phys. Rev.
{\bf C79}, 054004 (2009).
\bibitem{dong09} H. Dong, T.T.S. Kuo and R. Machleidt, Phys. Rev. {\bf C80},
065803 (2009).
\bibitem{nogga00} A. Nogga, H. Kamada and W. Glockle, Phys. Rev. Lett.
{\bf 94}, 944(2000).
\bibitem{pieper02} S.C. Peiper, K. Varga and R. B. Waringa, 
Phys. Rev. {\bf C66}, 044310 (2002).
\bibitem{navratil07} P. Navratil, V.G. Gueorguiev, J.P. Vary, W.E. Ormand
and A. Nogga, Phys. Rev. Lett. {\bf 99}, 042501(2007).
\bibitem{holt09} J. W. Holt, N. Kaiser and W. Weise, Phys. Rev. {\bf C79},
054331 (2009).
\bibitem{holt10} J. W. Holt, N. Kaiser and W. Weise, Phys. Rev. {\bf C81},
024002 (2010).
\bibitem{otsuka10} T. Otsuka, T. Sizuki, J.D. Holt, A. Schwenk and
Y. Akaishi, Phys. Rev. Lett. {\bf 105}, 032501 (2010).
\bibitem{idaho} D.\ R.\ Entem, R. Machleidt, and H. Witala, Phys.\ Rev.\ C 
{\bf 65}, 064005 (2002).
\bibitem{bogner01} S. K. Bogner, T. T. S. Kuo and L. Coraggio, Nucl. Phys.
{\bf A684}, (2001) 432.
\bibitem{bogner02} S.K. Bogner, T.T.S. Kuo, L. Coraggio
A.\ Covello, and N.\ Itaco, Phys.\ Rev.\ C {\bf 65}, 051301(R) (2002).
\bibitem{bogner03} S.\ K.\ Bogner, T.\ T.\ S.\ Kuo, and A.\ Schwenk,
Phys.\ Rep.\ {\bf 386}, 1 (2003).
\bibitem{ko90} T.\ T.\ S.\ Kuo and E.\ Osnes, {\it Lecture Notes in 
Physics} (Springer-Verlag, New York, 1990), Vol. 364.
\bibitem{klr} T.T.S. Kuo, S.Y. Lee and K.F. Ratcliff, {\em Nucl. Phys.}
 {\bf A176} (1971) 172.
\bibitem{lesu} S.Y. Lee and K. Suzuki, {\em Phys. lett.}
{\bf 91B}, 173 (1980).
\bibitem{sule} K. Suzuki and S.Y. Lee, {\em Prog. Theor. Phys.}
{\bf 64}, 2091 (1980).
\bibitem{okamoto10} R. Okamoto, K. Suzuki, H. Kumagai and S. Fujii,
to be published in `Proceedings of the 10th International Spring Seminar
on Nuclear Physics (May 21-25, Sur Mare Vietri, Italy, ed. by A. Covello)';
[nucl-th] arXiv:1011.1994v1.
\bibitem{suzu94}K. Suzuki, R. Okamoto, P.J. Ellis and T.T.S. Kuo, 
{\em Nucl. Phys.} {\bf A567} (1994) 576.
\bibitem{kren} E.M. Krenciglowa and T.T.S. Kuo, {\em Nucl. Phys.}
{\bf A235}, 171 (1974).
\bibitem{kuo95} T.T.S. Kuo, F. Krmpotic, K. Suzuki and R. Okamoto, 
{\em Nucl. Phys.} {\bf A582}, 205(1995).
\bibitem{dongzbox11} H. Dong, T.T.S. Kuo and J.W. Holt, preprint (May 2011,
 to be submitted to  arXiv and PRC).
\bibitem{song87} H. Q. Song, S. D. Yang and T.T.S. Kuo, Nucl. Phys. 
{\bf A462}, 491 (1987).
\bibitem{rentmeester} M. C. M. Rentmeester, R. G. E. Timmermans,
 and J. J. de Swart, Phys. Rev. C {\bf 67} (2003) 044001.
\bibitem{epel06} E. Epelbaum, Prog. Part. Nucl. Phys. {\bf 57}, 654(2006).



\bibitem{cdbonn} R.\ Machleidt, Phys.\ Rev.\ C {\bf 63}, 024001 (2001).

\bibitem{argonne} R.\ B.\ Wiringa, V. G. J. Stoks, and R. Schiavilla,
Phys.\ Rev.\ C {\bf 51}, 38 (1995).

\bibitem{nijmegen} V.\ G.\ J.\  Stoks, R. A. M. Klomp, C. P. F. Terheggen, 
and J. J. de Swart, Phys.\ Rev.\ C {\bf 49},
2950 (1994).

\bibitem{nucldata}http://www.nndc.bnl.gov/chart/.

\bibitem{ringschuck} P. Ring and P. Schuck, {\it The Nuclear Many-Body Problem}
(Springer-Verlag, New York, 1980).
\bibitem{bogner05} S. K. Bogner, A. Schwenk, R. J. Furnstahl and A. Nogga,
Nucl. Phys. {\bf A763} (2005) 59.
\bibitem{jdholt05} J. D. Holt, J.  W. Holt, T. T. S. Kuo, G. E. Brown and
S. K. Bogner, 
Phys. Rev. {\bf 72}, 041304R (2005).  


\end{thebibliography}
\end{document}